\documentclass[apl,article,twocolumn,superscriptaddress]{revtex4}

\usepackage{graphicx}
\usepackage{amssymb}
\usepackage{amsmath}
\usepackage{epsfig}
\usepackage{stmaryrd}
\usepackage[hypertex]{hyperref}

\newcommand{\un}   [1]{\ensuremath{\,\mathrm{#1}}}

\newcommand{\real}{\text{Re}}
\newcommand{\imag}{\text{Im}}

\newcommand{\tderl}[2]{\ensuremath{\text{d} #1/\text{d} #2}}

\begin{document}

\title{Discrete-time quadrature feedback cooling of a radio-frequency mechanical resonator}
\author{M. Poot\footnote{Present address: Department of Electrical Engineering, Yale University, New Haven, CT 06520,
USA}} \affiliation{Kavli Institute of Nanoscience, Delft
University of Technology, Lorentzweg 1, 2628 CJ Delft, The
Netherlands}
\author{S. Etaki}
\affiliation{Kavli Institute of Nanoscience, Delft University
of Technology, Lorentzweg 1, 2628 CJ Delft, The Netherlands}
\affiliation{NTT Basic Research Laboratories, NTT Corporation,
Atsugi-shi, Kanagawa 243-0198, Japan}
\author{H. Yamaguchi}
\affiliation{NTT Basic Research Laboratories, NTT Corporation,
Atsugi-shi, Kanagawa 243-0198, Japan}
\author{H. S. J. van der Zant}
\email{h.s.j.vanderzant@tudelft.nl} \affiliation{Kavli
Institute of Nanoscience, Delft University of Technology,
Lorentzweg 1, 2628 CJ Delft, The Netherlands}

\date{\today}

\begin{abstract}
We have employed a feedback cooling scheme, which combines
high-frequency mixing with digital signal processing. The
frequency and damping rate of a 2 MHz micromechanical resonator
embedded in a dc SQUID are adjusted with the feedback, and active
cooling to a temperature of 14.3 mK is demonstrated. This
technique can be applied to GHz resonators and allows for flexible
control strategies.
\end{abstract}

\maketitle
\newpage

Mechanical systems in the quantum regime
\cite{oconnell_nature_quantum_piezo_resonator,
teufel_arXiv_groundstate, poot_physrep_review} can be used to
answer fundamental questions about quantum measurement,
decoherence, and the validity of quantum mechanics in
macroscopic objects. This requires a mechanical resonator which
is cooled to such a low temperature that it is in its ground
state for most of the time. In the past few years tremendous
progress has been made in actively cooling resonators
\cite{poot_physrep_review}, mainly by using sideband cooling
\cite{arcizet_nature_cavity, rocheleau_nature_lown,
teufel_arXiv_groundstate} and active feedback cooling
\cite{kleckner_nature_feedback, poggio_PRL_feedback,
abbott_NJP_kg_groundstate}. The latter technique has mainly
been applied to low-frequency (kHz) resonators combined with
optical detection. The largest cooling factors have been
obtained using velocity-proportional feedback, i.e., by feeding
back the differentiated displacement signal. However, at higher
frequencies, delays in the feedback system seriously degrade
the cooling performance. Here, we demonstrate a feedback
cooling technique \cite{kriewall_JDS_heterodyne} with a nearly
unlimited bandwidth, based on fast digital signal processing
(DSP) in combination with single-sideband mixing. A 2 MHz
micromechanical resonator with inductive readout is cooled to
$14.3 \un{mK}$ using this scheme.

\begin{figure}[tbh]
\centering
\includegraphics[width=1\columnwidth]{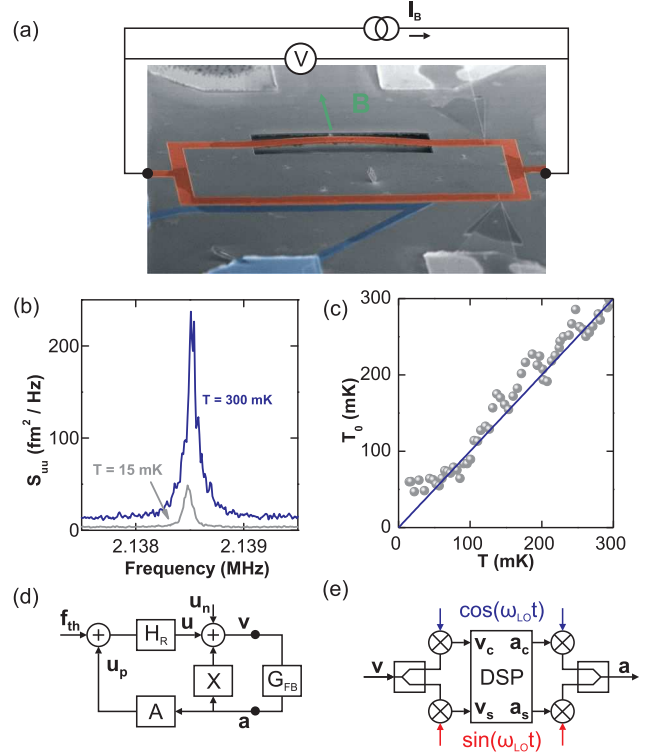}
\caption{(a) Schematic overview of the SQUID detector (red) with
the integrated flexural resonator. (b) Displacement noise spectra
without feedback. (c) The resonator temperature
extracted from the thermal noise spectra plotted against the
mixing chamber temperature. (d) Generic linear system
representation \cite{oppenheim_signals_systems} of feedback
cooling. (e) The feedback filter consists of a digital signal
processor with a single-sideband mixer at the input and output.}
\label{figure1}
\end{figure}

Figure \ref{figure1}a shows the device, which consists of a dc
SQUID with a part of its loop suspended. This forms a $50 \un{\mu
m}$ long flexural resonator with its fundamental mode around $f_0
\sim 2 \un{MHz}$. The chip is glued onto a piezo element for
feedback and actuation, and cooled in a dilution refrigerator with
a minimum bath temperature of $15 \un{mK}$. By applying an
in-plane magnetic field $B$ (green), a displacement of the beam
$u$ changes the amount of flux through the dc SQUID loop. When a
bias current $I_B$ is applied, a change in flux results in a
change in the SQUID voltage $V$. This way, the dc SQUID is a
sensitive displacement detector \cite{etaki_natphys_squid}. In all
measurements presented here, the same working point for the dc
SQUID is used, to avoid backaction-induced changes in frequency
and damping \cite{poot_PRL_backaction}.

The thermal noise of the resonator is used to calibrate the dc
SQUID detector. Figure \ref{figure1}b shows the displacement
noise spectrum $S_{uu}$ measured at two different cryostat
temperatures $T$. The thermal motion of the resonator shows up
as a peak on top of the imprecision noise floor $S_{u_n u_n}$.
The cryogenic-amplifier-limited displacement noise is
$2 \un{fm / \sqrt{Hz}}$ at $T = 15 \un{mK}$. The area under the
peak is the amplitude of the Brownian motion of the resonator
squared. When the temperature of the refrigerator is increased
to $T = 0.3 \un{K}$, the spectrum changes: Firstly, the noise
floor is higher due to a decrease in the transduction,
$\tderl{V}{u}$, as the critical current decreases with
increasing temperature \cite{etaki_natphys_squid}. Secondly,
the peak is higher and wider (the intrinsic damping rate
$\gamma_0$ increases with temperature), indicating that the
thermal motion is larger at higher temperatures. The resonator
temperature $T_0 = k_0 \langle u^2 \rangle / k_B$ ($k_0 = 110
\un{N/m}$ is the spring constant) is plotted against $T$ in
Fig. \ref{figure1}c: It follows the cryostat temperature for $T
> 50 \un{mK}$ (solid line) and saturates below this value.

To further lower the resonator temperature, active feedback is
employed, where the displacement of the resonator is fed back
to it to damp its thermal motion. Fig. \ref{figure1}d
illustrates the generic process \cite{poot_physrep_review}: The
thermal force noise $F_{th} \equiv k_0 f_{th}$ drives the
resonator whose response is $H_R = f_0^2/(f_0^2 - f^2 + i f
\gamma_0 /2\pi)$. Note, that $f_{th}$ and the other signals are
scaled to have the unit of position. The force results in a
displacement which is measured by the dc SQUID detector, and
imprecision noise $u_n$ is added to its output $v$. This signal
is fed to the feedback filter with transfer function $G_{fb}$.
The actuation $a$ is multiplied by $A$, which consists of the
SQUID transduction, an attenuation (-40 dB) and the piezo
responsivity. Finally, the resulting piezo displacement $u_p$
exerts an inertial force on the resonator. Note, that in
practise crosstalk ($X$) exists between the applied feedback
and the detector output, which modifies the system response.

To fully characterize the linear system an ac signal is applied
to $a$ (see Fig. \ref{figure1}d) and the response at $v$ is
measured at the same frequency, while sweeping the driving
frequency across the resonance. In this case, the feedback
$G_{fb}$ is disabled. From this network-analyzer measurements
the elements $A = 1.94 \cdot 10^{-4} \exp(-0.73i)$, and $X =
0.26\exp(2.56i)$ of the linear systems are obtained as well as
the parameters of $H_R$: $f_0$ and $\gamma_0$. The non-zero
phase of $A$ is due to the time it takes for the signal to
travel through the whole system. If an analog differentiator
would be used for $G_{fb}$, this delay causes the feedback to
not be purely velocity proportional thus degrading the cooling
performance. The DSP-based feedback presented here can
compensate for this effect as demonstrated below.

Our implementation of the feedback filter $G_{fb}$ is shown in
Fig. \ref{figure1}e. The high-frequency input signal $v$ is
split and both branches are mixed with local oscillator (LO)
signals with a $90^o$ phase difference between them. This IQ
mixer gives both quadratures $v_s$ and $v_c$ of the input
signal. The LO frequency is $f_{LO} = 2.0492 \un{MHz}$ so the
down-mixed signals oscillate at $f_R - f_{LO} = 8.9 \un{kHz}$.
They are digitized and the DSP (Adwin Pro II at a sampling rate
$f_s = 820 \un{kS/s}$) applies the following transformation to
the input signals to generate two output signals $a_c$ and
$a_s$:
\begin{equation}
\left(\begin{array}{c}
a_c \\
a_s
\end{array}\right) =
g_{fb} \left(\begin{array}{cc}
\cos \theta_{fb} & -\sin \theta_{fb}\\
\sin \theta_{fb} &  \cos \theta_{fb}
\end{array}\right)
\left(\begin{array}{c}
v_c \\
v_s
\end{array}\right)
\end{equation}
These quadratures are then up-converted by the LO frequency
with a second IQ mixer. The final result is a signal $a$ at the
original frequency that is phase-shifted by the feedback phase
$\theta_{fb}$ and multiplied by the feedback gain $g_{fb}$,
i.e. $G_{fb} = g_{fb} \exp (i \theta_{fb})$. The only frequency
requirement for this mixing scheme is that the quadratures do
not change faster than the sampling rate, which is equivalent
to $\gamma_R/2\pi \lesssim f_s/2$. The operation is thus not
limited to resonators with frequencies within the bandwidth of
the DSP, allowing feedback cooling of radio and microwave
frequency resonators. Note, that the imprecision noise floor is
still determined by the cryogenic amplifier; the contributions
from the mixers and discretization are negligible.

\begin{figure}[tbh]
\centering
\includegraphics[width=1\columnwidth]{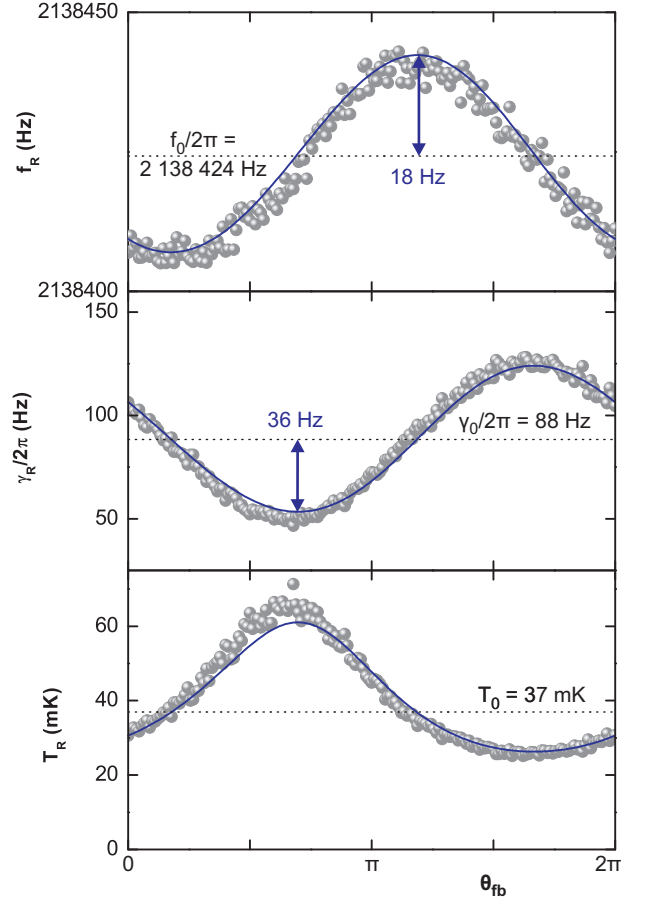}
\caption{Feedback phase dependence of the resonator frequency (a),
damping rate (b), and resonator temperature (c) for $g_{fb} = 0.1$. The dashed lines
indicate their measured values in the absence of feedback, ($f_0$, $\gamma_0$, and $T_0$ resp.); the
solid line is the phase dependence calculated using independent
measurements.} \label{figure2}
\end{figure}
The feedback modifies the resonator response from $H_R$ to its
closed-loop form $H_R'$ \cite{poot_physrep_review}:
\begin{equation}
H_R' = \frac{f_0^2}{f_0^2 - f^2 + i f \gamma_0 / 2 \pi - f_0^2 G_{fb}' A},
\end{equation}
where $G_{fb}' = g_{fb}' \exp(i \theta_{fb}') = G_{fb}/(1 - X
G_{fb})$ is the feedback filter modified by the crosstalk. The
real part of $G_{fb}' A$ modifies the resonance frequency from
$f_0$ to $f_R \approx f_0 (1 - \real[G_{fb}'A]/2)$, whereas the
imaginary part changes the damping from $\gamma_0$ to $\gamma_R
\approx \gamma_0 - 2 \pi f_0 \imag[G_{fb}'A]$. Both the
frequency shift and the change in damping depend periodically
on the phase of $G_{fb}'A$ and the maximum frequency shift is
half the maximum damping rate change.

In order to achieve optimal cooling the feedback phase is
varied for a fixed feedback gain as shown in Fig.
\ref{figure2}. The feedback gain is chosen sufficiently small
so that $G_{fb}' \approx G_{fb}$. At every point the thermal
noise spectra are measured and fitted to obtain the resonance
frequency (top), the damping rate (center), and the resonator
temperature (bottom). The resonance frequency and damping rate
show the expected sinusoidal dependence on the feedback phase.
The amplitude of the frequency shift is half of that of the
change in damping, consistent with the discussion above. The
phase where the damping is maximized, coincides with the lowest
resonator temperature and zero frequency shift. At this phase
the system delay is compensated and a pure
velocity-proportional feedback is applied to the resonator
(i.e. $\angle A G_{fb} = -\pi/2$). The phase dependencies can
also be calculated without any free parameters by using the
values from the network characterization (Fig. \ref{figure1}d).
Figure 2 show that these are in good agreement with the
feedback results.

\begin{figure}[tbh]
\centering
\includegraphics[width=1\columnwidth]{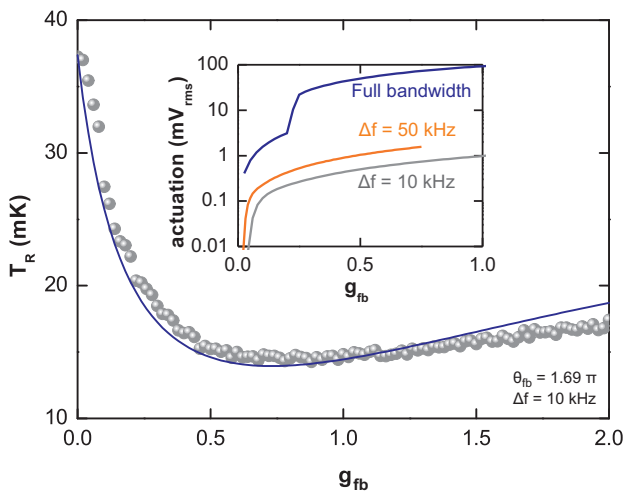}
\caption{Resonator temperature as a function of feedback gain, showing
both the feedback measurements (symbols) and calculations for pure
velocity-proportional feedback (solid line). The inset shows the root-mean-squared
value of the actuation signal ($\sqrt{\langle a_c^2 \rangle + \langle a_s^2 \rangle}$)
as a function of gain for different filter bandwidths.} \label{figure3}
\end{figure}
To further cool the resonator, the feedback gain is increased
at the optimal phase as indicated in Fig. \ref{figure3}. First
the resonator temperature decreases rapidly with increasing
gain due to the increased damping rate. However, by increasing
the gain further, more of the imprecision noise $u_n$ is fed
back as force noise. This causes a steady increase in $T_R$ for
large $g_{fb}$. The minimum temperature that can be reached is
set by $S_{u_n u_n}$ and the solid line shows the predicted
curve for velocity-proportional feedback
\cite{poggio_PRL_feedback, lee_PRL_feedback_toroid} calculated
with the experimental parameters. The achieved minimum of $14.3
\un{mK}$ is close to the predicted lowest temperature of $14.0
\un{mK}$. Note, that a temperature of $14.3 \un{mK}$
corresponds to an average thermal phonon occupation of $\bar n
\approx k_B T_R / h f_R = 138$ for a 2 MHz resonator. The
heterodyne DSP-based technique employed in this work thus
successfully reaches the lowest temperature possible for the
standard fully-analog approach, but now applied to a
high-frequency resonator.

Another advantage of DSP-based feedback is that the transformation
of $v_s$ and $v_c$ to $a_c$ and $a_s$ can be designed with almost
arbitrary transfer characteristics, allowing implementation of
optimal control strategies \cite{bruland_JAP_optimal_control}. In
the measurements in Fig. \ref{figure3}, the input signal is
digitally filtered using a Fourier transform filter which is
centered around $f_0$ and a tunable filter bandwidth $\Delta f$.
The filter reduces the bandwidth of the feedback which prevents
excess signal output outside the resonator bandwidth that may
overload the detector or the amplifiers. The inset of Fig.
\ref{figure3} shows the root-mean-square output voltage as a
function of $g_{fb}$ for three values of $\Delta f$. For the full
bandwidth ($f_s/2 = 410 \un{kHz}$) an instability occurs around
$g_{fb} = 0.23$, which affects the cooling. The 10 kHz bandwidth,
which is used for the cooling curve of Fig. \ref{figure3}, has two
orders of magnitude less actuation compared to the full bandwidth,
enabling efficient cooling without affecting the closed-loop
response as long as $\Delta f \gg \gamma_R/2\pi$. This again
illustrates the versatility of our digital quadrature feedback
cooling platform.

We thank Hidde Westra, Khasahir Babei Gavan, Abah Obinna and
Joris van der Spek for their help with the measurements. This
work was supported in part by FOM, NWO (VICI grant), NanoNed, a
EU FP7 STREP project (QNEMS), and JSPS KAKENHI (20246064 and
23241046).


\end{document}